\documentclass{ws-procs975x65}

\usepackage{latexsym}
\usepackage{graphicx}
\usepackage{pifont}

\newcommand{\req}[1]{Eq.\,(\ref{#1})}

\newcommand{\beqn}{\begin{equation}}
\newcommand{\eeqn}{\end{equation}}

\newcommand{\tr}{\mathrm{tr}\,}

\newcommand{\Veff}{\ensuremath{V_{\mathrm{eff}}}}
\newcommand{\cond}{\langle \bar\psi \psi \rangle}

\newcommand{\Fumn}{F^{\mu\nu}}

\newcommand{\calE}{{\cal E}}
\newcommand{\TEH}{T_{\rm M}}
\newcommand{\THU}{T_{\rm U}}
\newcommand{\fracg}{\frac{g}{2}}

\begin{document}

\title{Temperature of Electron Fluctuations in an Accelerated Vacuum}

\author{Lance Labun$^*$ and Johann Rafelski}

\address{Department of Physics, The University of Arizona,
Tucson, AZ 85721, USA\\
$^*$E-mail: lalabun@labunllc.com}

\begin{abstract}
The electron vacuum fluctuations measured by $\cond$ do not vanish in an externally applied electric field $\calE$. For an exactly constant field, that is for vacuum fluctuations in presence of a constant accelerating force, we show that $\cond$ has a Boson-like structure with spectral state density $\tanh^{-1}(E/m)$ and temperature $\TEH = e\calE/m\pi = a_v/\pi$. Considering the vacuum fluctuations of `classical' gyromagnetic ratio $g=1$ particles we find Fermi-like structure with the same spectral state  density at a smaller temperature $T_1=a_v/2\pi$ which corresponds to the Unruh temperature of an accelerated observer.
\end{abstract}


\bodymatter

\section{Introduction}
M\"uller et al.~\cite{Muller:1977mm} showed that in presence of constant external electromagnetic fields the Heisenberg-Euler effective potential of QED can be cast in the form of a thermal background characterized by a spectral density of states $\rho(E)$ and temperature $\TEH$
\beqn\label{TEH}%\label{MulSpectra}
\rho(E)= \frac{m^2}{8\pi^2}\ln(E^2-m^2+i\epsilon),\qquad
\TEH=\frac{e\calE}{m\pi}=\frac{a_v}{\pi}\,.
\eeqn
Since an electric field accelerates all charged particles, real or virtual, this can be understood as a property of the vacuum under constant global acceleration~\cite{Greiner:1985ce} $a_v=e\calE/m$.  This circumstance has also been discussed by Pauchy Hwang and Kim~\cite{PauchyHwang:2009rz}.

Associated with this result is the quantum statistics: spin-1/2 QED exhibits a thermal distribution as though the fluctuating degrees of freedom are bosons, while spin-0 scalar QED exhibits vacuum fluctuations as though the degrees of freedom are fermionic. The former result could be understood as being due to pairing of electron-positron pairs.  However the latter case has no obvious explanation.  

The case of `accelerated vacuum'  parallels  that of an `accelerated observer' traveling in a matter- and field-free spacetime. This observer sees a thermal background characterized by the Unruh temperature
\beqn\label{TUH}
\THU=\frac{a_{\rm U}}{2\pi}=\frac{\TEH}{2}.
\eeqn
The statistics of the thermal distribution are bosonic considering the vacuum of a scalar particle~\cite{Unruh:1976db,Crispino:2007eb} and fermionic in the vacuum of a fermi particle~\cite{Candelas:1978gg}. Many  readers  will expect that there should not be a difference between `accelerated vacuum' and  `accelerated observer', yet there is a disagreement in relation to particle spin and statistics and  the value of the temperatures \req{TEH} \& \req{TUH}.

We will show that these two results \req{TEH} \& \req{TUH} are experimentally distinguishable. This implies the ability to determine which is accelerated, the observer or the vacuum state. The question is whether or not one should be able in-principle to determine the accelerated state within the current formulation of quantum field theory. If the conclusion is that one should {\it not} be able to determine which is accelerated, then there is additional undiscovered physics content in either the vacuum structure or the Unruh accelerated detector.

It is of considerable interest to find a model achieving agreement in the quasi-thermal properties of the quantum vacuum, in the sense that the accelerated observer registers the same outcome as the `accelerated' field-filled vacuum.  The disagreement in relation to particle spin suggests a  closer look at the spin properties of the fluctuations. Altering the magnetic moment of the electron from its Dirac value described by the gyromagnetic ratio  $g=2$ to the `classical' spinning-particle value $g=1$ achieves agreement with the Unruh temperature and statistics~\cite{Labun:2012jf}.  We summarize the results obtained and consider the properties of QED vacuum condensate  $\cond$ in presence of strong quasi-constant external fields in the thermal framework modeled by the `classical' $g=1$ QED.

%%%%%%%%%%%%%%%%%%%%%%%%%%%%%%%%%%%%%%%%%%%%%%%
\section{Effective Action}
Charge convective current and spin magnetic dipole current are conserved independently and thus in QED the associated integral `charges' - the electric charge and magnetic dipole moment can be prescribed arbitrarily. The gyromagnetic ratio $g$ combines with particle charge  and mass in defining the magnitude of the magnetic dipole moment. The dynamics of a particle $\psi$ with general $g$ is generated by
\beqn\label{EoMg}
\left[D^2+m^2-\frac{g}{2}\frac{e\sigma_{\mu\nu} \Fumn}{2}\right]\psi=0
\eeqn
where $D=i\partial+eA$ is the covariant derivative, $\Fumn$ the electromagnetic field strength tensor and $\sigma_{\mu\nu}=(i/2)[\gamma_{\mu},\gamma_{\nu}]$. Only for the specific case $g=2$ can one choose to write this in  the Dirac equation form. Any value of the gyromagnetic ratio  $g$ can arise, and as long as the quantization of charge is not understood, it is difficult to claim that the value of $g$ arising in a specific simplified dynamical equation is of greater interest than other values. Like charge, $g$ is the subject of experimental and theoretical effort to determine quantum corrections to an input value, the best known being the QED correction to the Dirac particle $g=2+\alpha/\pi+\ldots$.

The effect of $g$ on vacuum fluctuations is determined from the  effective potential
\beqn\label{Veffdefn}
\Veff=-\frac{i}{2}\tr\ln\left[D^2+m^2-\frac{g}{2}\frac{e\sigma_{\mu\nu} \Fumn}{2}\right]
\eeqn
For the Heisenberg-Euler case of a constant electric-only field of strength $\calE$, we evaluate the trace using the proper time method~\cite{Schwinger:1951nm}
\beqn\label{Veff}
\Veff = \frac{\gamma_s}{32\pi^2}\!\!\int_{0}^{\infty}\!\frac{du}{u^{3}}\left(\frac{e\calE u\cosh(\frac{g}{2}e\calE u)}{\sinh e\calE u}-1\right)e^{-im^2u}
\eeqn
in which  the generalized degeneracy $\gamma_s$ counts the number and type of degrees of freedom.  $\gamma_s=4$ when $g=2$ for spin-1/2 Dirac electron, or $\gamma_s=-2$ when $g=0$ for a spin-0 electron.  The $-1$ inside the parentheses removes the field-independent constant. 

Transforming $\Veff$ to a statistical format proceeds via meromorphic expansion of the integrand of \req{Veff}~\cite{Muller:1977mm,Labun:2012jf}.  The finite (regularized and renormalized) effective potential is
\beqn\label{Veff-step1}
\Veff  = 
\frac{\gamma_sm^2\TEH^2}{32\pi^2}\!\int_0^{\infty}\!\!\frac{2u\:du}{u^2\!-\!1\!+\!i\epsilon}\sum_{n=1}^{\infty}
\frac{e^{-nu\frac{m}{\TEH}}}{n^2}\cos\left(n\pi(\fracg+1)\right)
\eeqn
with integration contour defined by the usual assignment
\beqn\label{msqminusie}
m^2\to m^2-i\epsilon.
\eeqn
This also makes explicit that the effective potential contains an imaginary part, as will be discussed below.  Setting $g=2$ for a spin-1/2 (Dirac) electron, $\cos 2n\pi = 1$ for all $n$, and setting $g=0$ for a spin-0 electron, $\cos n\pi = (-1)^n$ producing an alternating sum.  For each case, integrating by parts twice and summing the series yields the results of M\"uller et al~\cite{Muller:1977mm}.

The exponential weights of the terms in the series in \req{Veff-step1} generate a thermal distribution, and the statistics of the distribution are determined by the phase of the terms in the series.  
Summing with arbitrary $g$, the effective potential is
\beqn\label{Veffggen}
\Veff = \frac{\gamma_sm^2\TEH}{64\pi^2}\!\int_0^{\infty}\!\!\!dE \ln(E^2\!-\!m^2\!+\!i\epsilon)\!\sum_{\pm}\ln(1+e^{\pm i\pi\fracg}e^{-E/\TEH})
\eeqn
The sum over $\pm$ ensures the distribution is real so that the imaginary part arises only from the branch cut in the first log factor. Restoring $g=2$ and $\gamma_s=4$ identifies the spectral density of states \req{TEH} according to $V\equiv T\int_0^{\infty}\ln(1-e^{-E/T})\rho(E)dE$.

For $g=1$ summing over $\pm$ simplifies it to
\beqn\label{Veffg1}
\Veff\Big|_{g=1} \!\!\!= \frac{\gamma_sm^2\THU}{32\pi^2}\!\!\int_0^{\infty}\!\!\!\!dE \ln(E^2\!-\!m^2\!+\!i\epsilon)\ln(1+e^{-E/\THU})
\eeqn
exhibiting in the second log factor a thermal fermionic distribution controlled by the Unruh temperature, $\THU$.  The effective potential of a `classical spinning electron' with $g=1$ in a constant field thus has the format of a thermodynamic potential with temperature parameter and statistics in agreement with the expectation of an accelerated observer in the (unaccelerated) vacuum of a fermion field.   

%%%%%%%%%%%%%%%%%%%%%%%%%%%%%%%%%%%%%%%%%%%%%%%
\section{Condensate}
The quantum fluctuations induced by the external field are measured by the condensate $\cond$, which is the difference of the Green's functions in the vacuum with the external field and the vacuum with no field present (0 superscript)
\beqn\label{conddefn}
-\cond=\tr\left[iS_F(x,x)-iS_F^0(x,x)\right], \quad S_F(x,x')=-i\langle \mathcal{T}\psi(x')\bar\psi(x)\rangle\,.
\eeqn
$\langle\mathcal{T}...\rangle$ is the vacuum expectation of time-ordered operators, the limit $x'\to x$ is evaluated in the point-splitting procedure to preserve gauge invariance, and the $F$ subscript indicates Feynman boundary conditions.  This definition of the condensate displays the implicit definition of the reference, no-field vacuum state.

The condensate is related to the effective potential by
%An important measure of the quantum fluctuations is the condensate $\cond$, which is related to the effective potential by
\beqn\label{condVeffreln}
-m\langle\bar\psi\psi\rangle=m\frac{d\Veff}{dm}
\eeqn
Note that the differentiation with respect to $m$ improves the convergence properties of \req{Veff}.
Evaluating the derivative of \req{Veff}, we find the formerly logarithmically divergent contribution is finite.  Because $-m\cond$ includes this term quadratic in $\calE$ (see discussion in~\cite{Labun:2008qq}), we use now the meromorphic series 
\beqn \label{meroexp2} 
1-\frac{x\,\cosh(xy)}{\sinh (x)}=-2x^2\sum_{n=1}\frac{\cos n\pi (y+1)}{x^2+(n\pi)^2}
\eeqn
The resulting expression for $\cond$ for arbitrary $g$ is
\beqn\label{condggen}
%\cond = -\frac{\gamma_sm^2}{32\pi^2}\!\sum_{\pm}\int_0^{\infty}\!\!\!\!dE \frac{\tanh^{-1}(E/m+\!i\epsilon)}{e^{E/\TEH}+e^{\pm i\pi\fracg}}e^{\pm i\pi\fracg}
\cond = \frac{\!-\gamma_sm^2\!}{8\pi^2}\!\int_0^{\infty}\!\!\!\!dE \frac{\tanh^{-1}(E/m+\!i\epsilon)(1\!+\!e^{E/\TEH}\cos(\fracg\pi))}{e^{2E/\TEH}+1+2e^{E/\TEH}\cos(\fracg \pi)}
\eeqn

Setting $g=0$ ($g=2$) yields a fermionic (bosonic) distribution controlled by $\TEH$ 
\begin{align}
\cond=-\frac{m^2}{4\pi}\frac{\gamma_s}{2\pi}\int_0^{\infty}\frac{\tanh^{-1}(E/m+i\epsilon)dE}{e^{E/\TEH}+(-1)^{g/2}}, \label{condg02}
\end{align}
Identifying the numerator of the integrand $\tanh^{-1}(z)$ by analogy with statistical physics $\langle N\rangle/V=\int_0^{\infty}\Gamma(E)dE/(e^{E/T}\pm 1)$, the degeneracy of states 
\beqn
\Gamma(E)=-\frac{\gamma_s}{4}m^2\tanh^{-1}(E/m)
\eeqn 
is the same in each value of $g$ considered. 

For $g=1$, the terms containing cosine in the numerator and denominator of \req{condggen} vanish, leaving the fermi occupancy factor in the denominator with twice the Euler-Heisenberg temperature $2/\TEH=1/\THU$.  
\beqn\label{condg1}
\cond_{g=1}=\frac{m^2}{4\pi}\frac{\gamma_s}{2\pi}\int_0^{\infty}\frac{\tanh^{-1}(E/m+i\epsilon)dE}{e^{E/\THU}+1}
%\sum_{n=1}\frac{(-1)^n}{2n}e^{-2n\beta mu/2}=-\frac{1}{2}\ln(1+e^{-\beta mu})
\eeqn
which displays the fermionic occupancy factor in the denominator with a distribution controlled by the Unruh temperature. The fluctuations of a $g=1$ `electron' in an external electric field are thus reconciled with the fluctuations expected by an observer accelerated at $a_v=e\calE/\pi m$ through the (unaccelerated) vacuum of a fermion field.   

%%%%%%%%%%%%%%%%%%%%%%%%%%%%%%%%%%%%%%%%%%%%%%%%%%%%%
\section{Observable Effects from $g=1$}
One observable effect is spontaneous pair  production in strong fields discussed by us earlier~\cite{Labun:2012jf}, and originating in the imaginary part of the effective action \req{Veff-step1}.  This allows any non-vanishing electric field to spontaneously decay into charged particle pairs. $g=0,2$ yield the largest total decay probability and are in this sense the best cases for experiment.  For $g=1$ the reduction in the effective temperature parameter is largest.  Due to the exponential dependence, the reduction in the temperature parameter by factor 2 reduces spontaneous pair production below the critical field $\calE_c=m^2/e$ by many orders of magnitude.

The real part of the effective potential generates nonlinear field-field interactions.  These interactions are exhibited order-by-order in the field by expanding $\Veff$ in a (semi-convergent) power series in $(e\calE)^{2n}$.
The power series representation of $\Veff$ is obtained by expanding the proper time integrand of \req{Veff} for $e\calE u\ll 1$
\begin{align}
\Veff \simeq &\frac{\gamma_s}{32\pi^2}\left\{
\left(7-\frac{15}{2}g^2+\frac{45}{48}g^4\right)\frac{1}{45}\frac{(e\calE)^4}{m^4}\right.\\ \notag &+ \left(\frac{31}{24}-\frac{49}{32}g^2+\frac{35}{128}g^4-\frac{7}{512}g^6\right)\frac{4}{315}\frac{(e\calE)^6}{m^8}+...
\end{align}
At each order in $(e\calE)^{2}$, we have separated the numerical coefficients for $g=2$ outside the parentheses for ease of comparison to the $g=1$ result,
\beqn
\Veff\Big|_{g=1}\!\!\!\simeq \frac{\gamma_s}{32\pi^2}\left\{\frac{7}{5760}\frac{(e\calE)^4}{m^4}-\frac{31}{161280}\frac{(e\calE)^6}{m^8}...\right.
\eeqn
For example, the ratio of the coefficients of the $(e\calE)^4$ terms is $\Veff(g=1)/\Veff(g=2)\simeq 7/128$.
We see that nonlinear field-field interactions generated by this potential are suppressed in the $g=1$ case relative to the $g=2$ (or $g=0$) electron.  This outcome is consistent with the suppression of the imaginary part of $\Veff$.  Experiments seeking nonlinear field effects~\cite{Rikken:2001zz,QAexperiment} derived from the Euler-Heisenberg effective potential are thus also sensitive to the effective value of $g$.

%%%%%%%%%%%%%%%%%%%%%%%%%%%%%%%%%%%%%%%%%%%%%%%%%%%%%%
\section{Discussion and conclusions}
In summary, we have recalled that in a constant electric field $\calE$, the electron fluctuations $\langle \bar\psi\psi\rangle$ display a thermal Bose spectrum with temperature $\TEH=e\calE/m\pi=a_v/\pi$ \req{condg02}.  This result contrasts with the Fermi spectrum and Unruh temperature $\THU=a_{\rm U}/2\pi$ expected from viewing the vacuum fluctuations of the electrons as accelerated.  We have calculated $\cond$ in an electric field for the gyromagnetic ratio $g=0,1,2$.  Setting $g=1$, as though considering the quantum fluctuations of a `classical spinning particle', displays the Unruh $\THU=a/2\pi$ and a Fermi spectrum, see \req{condg1}. 

We highlight the functional dependence of light-light scattering on $g$ because it has implications for future experiments.  Any (effective) value of $g$ deviating from the Dirac value $g=2$ results in a suppression of the rate of light-light scattering.  We note that QED is not yet tested near the critical field strength $\calE_c=m^2/e$, and in this strong-field regime, we have still to validate the approach to calculating Eq.\:\eqref{Veff}, which is perturbative in $\alpha$~\cite{Heisenberg:1935qt}.  Even more to the point there are serious questions about validity of QED in this limit~\cite{Rafelski:LeCosPA}. Therefore, the connection which we established to reconcile the two ways of viewing acceleration could forebear forthcoming theoretical developments.

An observable, physical difference such as now predicted in Eqs.~\eqref{TUH} and \eqref{TEH} provides an in-principle test to determine whether the strong field theory is valid, and/or it is the observer or the vacuum that is accelerated.  Being able to determine who is accelerated means a fixed reference frame has been selected and defined as inertial.  In quantum theory, the quantum vacuum state is a natural candidate for the fixed reference frame~\cite{Rafelski:LeCosPA}, and here we have recalled that the electron condensate contains in its definition \req{conddefn} a specific vacuum state as reference.  Experimental observation of quantum vacuum phenomena such as light-light scattering offers an important test of our understanding of the vacuum state canonically selected in quantum field theory and may reveal whether or not it is consistent with the vacuum selected in the Unruh accelerated detector situation.\\[-3mm]

%%%%%%%%%%%%%%%%%%%%%%%%%%%%%%%%%%%%%%%%%%%%%%%%%%%%%%%%%%%%%%%%
{\it Acknowledgments:}  L.L. thanks Director Pisin Chen for the opportunity to visit LeCosPA.  This work was supported by a grant from the  US Department of Energy, DE-FG02-04ER41318. \\[-8mm]
%%%%%%%%%%%%%%%%%%%%%%%%%%%%%%%%%%%%%%%%%%%%%%%%%%%%%%%%%%%%%%%%

\end{document}